\documentclass[sigconf]{acmart}

\usepackage{colortbl}      
\usepackage[table]{xcolor}
\usepackage{amsmath}
\usepackage{latexsym}
\usepackage{multirow}
\usepackage{enumitem}
\usepackage{makecell}
\usepackage{mdframed} 
\usepackage{ragged2e} 
\usepackage{algorithm}
\usepackage{algpseudocode}
\usepackage{arydshln}
\usepackage{enumitem}

\usepackage[table,dvipsnames]{xcolor}

\newmdenv[
  linewidth=0.8pt,
  roundcorner=6pt,
  backgroundcolor=gray!8,
  linecolor=black,
  innerleftmargin=10pt,
  innerrightmargin=10pt,
  innertopmargin=8pt,
  innerbottommargin=8pt,
  skipabove=10pt,
  skipbelow=10pt
]{promptbox}

\newcommand{\miniskip}{\vspace*{-.5\baselineskip}}

\newcommand{\proposedmethod}{LeMUQ}
\newcommand{\MainTask}{Uncertainty Quantification}
\newcommand{\MainTaskAbbr}{UQ}
\newcommand{\LARSRet}{$\text{LARS}_{Ret}$}
\newcommand{\LARSBase}{$\text{LARS}_{Base}$}

\newcommand{\query}{q}
\newcommand{\doc}{c}
\newcommand{\image}{I}
\newcommand{\response}{r}

\AtBeginDocument{%
  }

\setcopyright{acmlicensed}
\copyrightyear{2018}
\acmYear{2018}
\acmDOI{XXXXXXX.XXXXXXX}
\acmConference[Conference acronym 'XX]{Make sure to enter the correct
  conference title from your rights confirmation email}{June 03--05,
  2018}{Woodstock, NY}
\acmISBN{978-1-4503-XXXX-X/2018/06}

\begin{document}

\title{Uncertainty Quantification for Multimodal Retrieval Augmented Generation}

\author{Simon Binz}
 \affiliation{
   \institution{Radboud University}
   \city{Nijmegen}
   \country{The Netherlands}}
 \email{simon.binz@ru.nl}
\additionalaffiliation{
  \institution{German Research Center for Artificial Intelligence (DFKI)}
  \city{Osnabrück}
  \country{Germany}
}

\author{Heydar Soudani}
\affiliation{%
  \institution{Radboud University}
  \city{Nijmegen}
  \country{The Netherlands}}
\email{heydar.soudani@ru.nl}

\author{Faegheh Hasibi}
\affiliation{%
 \institution{Radboud University}
 \city{Nijmegen}
 \country{The Netherlands}}
\email{faegheh.hasibi@ru.nl}

\renewcommand{\shortauthors}{Binz et al.}

\begin{abstract}
Retrieval Augmented Generation (RAG) improves the question answering capabilities of Large Language Models (LLMs) by incorporating external knowledge and has recently been extended to multimodal settings through Vision-Language Models (VLMs) that integrate visual and textual information.
Despite these advances, generated answers can still be incorrect or misleading. \MainTask\ (\MainTaskAbbr) methods aim to estimate the reliability of model outputs, but most existing approaches are designed for text-only models and perform poorly in multimodal RAG scenarios. A key challenge is capturing uncertainty arising from multiple stages of the pipeline, including retrieval, visual understanding, and generation. In this work, we show that modeling uncertainty using multimodal and retrieval-aware probability signals improves estimation in multimodal RAG systems. We introduce LeMUQ, a Learnable Multimodal UQ method that analyzes token probabilities under input modifications, such as removing modalities or retrieved context. By encoding these signals as probability tokens and processing them with a finetuned model, our approach captures interactions between modalities and retrieval. Experiments across datasets, retrievers, and VLMs show consistent improvements over baseline and finetuned \MainTaskAbbr\ methods. Our proposed \proposedmethod\ increases the AUROC metric by $3.8\%$ on average. Additionally, our method shows strong generalization performance across different retrieval setups and datasets with mixed results when transferring across different VLMs.
Our findings highlight the importance of modeling multimodal uncertainty and provide a step toward more reliable and safer multimodal RAG systems. 
Code is available on GitHub.\footnote{\href{https://github.com/uqmultimodalrag2026-beep/UQformultimodalRAG}{https://github.com/uqmultimodalrag2026-beep/UQformultimodalRAG}}
\end{abstract}




\maketitle

\section{Introduction}~\label{sec:intro}

\begin{figure*}[t]
\centering
\includegraphics[width=0.99\textwidth]{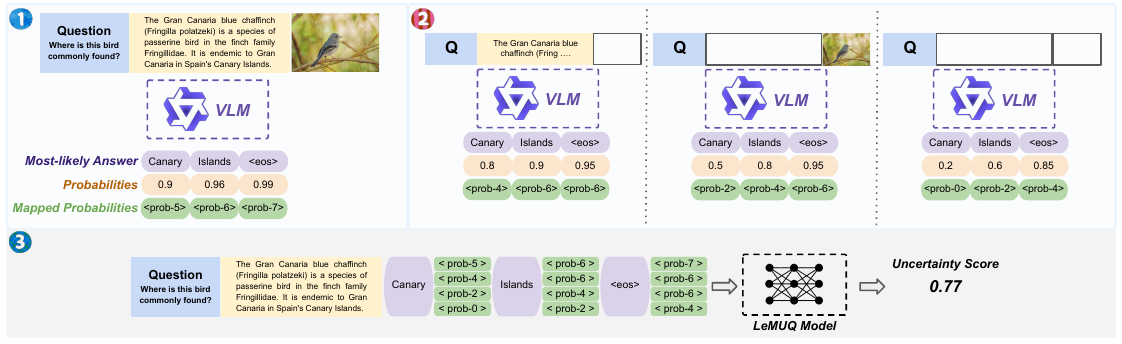}
\caption{Overview of the LeMUQ pipeline. 
\textbf{(1)} A VLM receives the query, input image, and retrieved context, and generates a response as a sequence of tokens along with their token probabilities, which are discretized into quantile bins and represented as mapped token probabilities.
\textbf{(2)} To account for different sources of uncertainty, the response is re-evaluated under multiple conditioning settings (without image, without retrieved context, and without both), producing alternative token probability sequences that are similarly converted into mapped probabilities.
\textbf{(3)} The query, retrieved context, generated response tokens, and corresponding mapped token probabilities from all configurations are concatenated and passed to a finetuned RoBERTa-based LeMUQ model, which outputs a scalar score estimating the VLM’s confidence in the generated response.
}
\label{fig:LeMUQ}
\end{figure*}

Large Language Models (LLMs) have shown strong performance across a wide range of tasks~\cite{surveyllm, Soudani2026, lin2025bagoftext, joko2026face}, such as question answering (QA), where queries are in natural language~\cite{qasurvey, Hoveyda2026orlog}. However, many QA tasks require specific knowledge that is not fully captured during the training stage of LLMs~\cite{mallen23popqa, Soudani24FTvsRAG, Soudani25Enhancing}. Retrieval-Augmented Generation (RAG) has been proposed to address this issue by incorporating externally retrieved knowledge, thereby improving answer quality~\cite{rag, ragknowledge, press2023selfask, Jin25SearchR1, rafiee2026trqa}.
As tasks and models grow more complex, multimodal extensions such as Vision-Language Models (VLMs) have been developed for settings like Visual Question Answering (VQA), where questions are paired with images~\cite{llava, li2023blip, LiuXL0023Universal}. These images provide important visual cues, motivating extensions of standard RAG setups to multimodal scenarios that integrate visual information~\cite{yu2025visrag, multimodalrag}.

Despite the improvements introduced by RAG systems and multimodal extensions, generated answers can still be incorrect or misleading~\cite{joren2025sufficient, Soudani24FTvsRAG, mallen23popqa}. This is particularly critical in domains such as medical or legal applications, where incorrect responses can have severe consequences~\cite{bakman2024mars, raghealth, raglegal}. To mitigate this issue, \textbf{Uncertainty Quantification (UQ)} methods aim to estimate the correctness of a model’s response, where high uncertainty indicates a likely incorrect response and low uncertainty suggests correctness~\cite{Zhao24Knowing, Hou24Decomposing, uqintroduction, Kadavath2022LanguageM, Harrison24Variational, Yin24Reasoning}. 

\smallskip\noindent\textbf{Research gap.} Existing research on UQ is mainly designed for text-only LLMs~\cite{malinin2021uncertainty, bakman2024mars, Soudani25Enhancing, lin2024generating, Feng24Hallucinate, MadhusudhanMYH25, Uncertainty25Heydar, Ekaterina26Throw, vazhentsev2025attention}, with a limited number of approaches measuring uncertainty in VLMs~\cite{khan2024consistency, ChaeK22Visual}. 
These standard \MainTaskAbbr\ methods, even those designed for VLMs, often achieve suboptimal performance in multimodal RAG settings, where extra (non-)textual information is introduced~\cite{xuan2025seeing}. Prior work has shown that many \MainTaskAbbr\ methods degrade in performance when non-parametric knowledge is incorporated~\cite{soudani2025uncertainty, Uncertainty25Heydar, Zhao25Propagation, perez2025uncertainty}. This is particularly problematic in multimodal RAG, where the input prompt is not purely textual and retrieved supporting context can increase model confidence in ways that do not necessarily align with correctness, thereby degrading calibration and increasing false-positive confidence~\cite{qiu2025entropy}.


\smallskip\noindent\textbf{Challenge.} In multimodal RAG settings, uncertainty can arise at different steps of the generation pipeline~\cite{yu2025visrag}. Within a single modality, inputs such as images or text may be ambiguous or noisy~\cite{khan2024consistency, joren2025sufficient}. Additional uncertainty may be introduced during the retrieval process, whether through text-only retrieval or multimodal retrieval pipelines such as image captioning followed by retrieval~\cite{yu2025visrag}. Further uncertainty can arise during the aggregation of information across modalities, as well as during the generation process itself~\cite{li2023blip, xuan2025seeing}. A UQ model for multimodal RAG needs to account for these sources of uncertainty and combine them in a calibrated manner to produce reliable uncertainty scores.  

\smallskip\noindent\textbf{Method.} In this work, we introduce a \textbf{Le}arnable \textbf{M}ultimodal \textbf{U}n\-cer\-tainty \textbf{Q}uantification method (\textbf{\proposedmethod}) that incorporates both multimodal and retrieval-based uncertainty. Our approach generates responses under varying conditions, including settings where specific modalities or retrieved context are selectively removed. The resulting response probabilities are encoded as special probability tokens and processed by a finetuned scoring model, which predicts an uncertainty score for each response. 

\smallskip\noindent\textbf{Experiments.} We conduct our experiments across multiple datasets, retrievers, and VLMs to evaluate the effectiveness of our proposed method in multimodal RAG settings. We first assess performance in a within-distribution setting, where the same retriever, dataset, and VLM are used during training and evaluation. Our results show that \proposedmethod\ consistently outperforms both baseline and finetuned \MainTaskAbbr\ methods, demonstrating the benefits of incorporating information about different modalities. In the in-distribution setting, our method achieves an average AUROC of 0.88, improving AUROC by 3.8\% on average across two datasets, different VLMs, and retriever setups.  


\medskip\noindent\textbf{Generalizability.} Previous research has shown that supervised methods often generalize poorly when the underlying data distribution shifts~\cite{distshift}. This limitation arises because such models tend to overfit to the specific characteristics of the training distribution, which leads to decreased performance on different validation and test distributions. To examine the generalizability of our method,  we evaluate LeMUQ under out-of-distribution settings across three dimensions: varying the retriever, the dataset, and the VLM. Our findings indicate that \proposedmethod\ maintains strong performance relative to the baseline and other finetuned methods. The performance gain is largely preserved when generalizing across different retrievers, generally surpasses both finetuned and baseline methods in dataset-transfer settings, and yields mixed results when transferring across different VLMs.

\medskip\noindent\textbf{The main contributions} of this paper are:
\begin{enumerate}
\item We investigate  the problem of UQ for multimodal RAG for the first time, to the best of our knowledge, and demonstrate the limitations of existing \MainTaskAbbr\ methods when applied to a multimodal RAG setting.
\item We propose \proposedmethod, a \MainTaskAbbr\ method for multimodal RAG that leverages probability signals under different input configurations to capture uncertainty arising from both retrieval and multimodal inputs.
\item We conduct experiments across multiple datasets, retrievers, and VLMs, demonstrating that our method consistently improves uncertainty estimation and generalizes well across different settings.
\end{enumerate}


\section{Related Work}~\label{sec:related_work}

\noindent
\textbf{Multimodal RAG.}
Multimodal RAG extends the text-only pipelines of standard RAG by incorporating multiple modalities such as images or audio into both the retrieval and generation stages, enabling access to richer information sources and potentially improving answer quality~\cite{multimodalrag}. VisRAG~\cite{yu2025visrag} proposes a retrieval framework that operates on document pages as unified multimodal units, allowing the model to reason over jointly embedded visual and textual information rather than treating modalities separately.
%
EchoSight~\cite{yan2024echosight} augments VLMs with external knowledge from Wikipedia articles through a two-stage retrieval process, first performing visual-only retrieval followed by multimodal reranking to identify relevant information. ReflectiVA~\cite{cocchi2025augmenting} introduces reflective tokens in a multimodal LLM that enable the model to decide whether retrieval is necessary and to assess the relevance of retrieved information during generation. In this paper, we focus on a standard multimodal RAG setup in which the context, image, and user query are passed to a generative VLM.

\noindent
\textbf{\MainTask\ for VLMs.}
In classical machine learning, uncertainty is typically characterized as a property of a model’s predictive distribution given a particular input~\cite{lin2024generating, liu2025uncertainty}. Building on this notion, \MainTaskAbbr\ aims to quantify the degree of variability or unpredictability in a model’s outputs, independent of any single generated response~\cite{kendall2017uncertainties, gawlikowski2023survey}.
Most existing approaches to \MainTaskAbbr\ have been developed in the context of question answering, where the LLM is treated as the sole source of uncertainty~{\cite{malinin2021uncertainty, lin2024generating}. More recently, however, this perspective has been extended to RAG, where uncertainty arises not only from the generative model but also from the retriever. In this setting, \MainTaskAbbr\ has been studied by modeling the relationship between retrieved documents and generated responses~\cite{soudani2025uncertainty}, utility-based models~\cite{perez2025uncertainty}, and fact-checking~\cite{fadeeva2025faithfulness}.

When moving to multimodal systems, additional sources of uncertainty emerge, particularly from visual inputs~\cite{Uncertainty25Heydar, fadeeva2025faithfulness}. Despite this, only a limited number of studies have explored \MainTaskAbbr\ for VLMs. For instance, ReCoVERR~\cite{srinivasan2024selective} estimates uncertainty via self-evaluation, prompting the model to assess the reliability of its own outputs. In contrast, HARMONY~\cite{harmony} combines multiple signals, including generated tokens, output probabilities, and the model’s internal assessment of visual understanding, to derive an uncertainty score. It compares the token distribution generated from the original image with that obtained after removing question-relevant visual evidence. The discrepancy between these distributions serves as an uncertainty measure, motivated by the intuition that unreliable predictions are less sensitive to such perturbations. ~\citet{avestimehr2025detecting} estimate uncertainty by comparing a model’s output distributions for an original image and a perturbed version where question-relevant visual evidence is removed. If the distributions remain similar despite removing key visual information, the response is likely unreliable.
Nevertheless, existing work has not yet addressed \MainTaskAbbr\ in the context of multimodal RAG, where uncertainty originates from three distinct components: the retriever, the generator, and the visual input. To bridge this gap, we propose a novel \MainTaskAbbr\ framework that estimates uncertainty by analyzing model probabilities under diverse input configurations.

\section{Preliminaries}~\label{sec:preliminaries}

\noindent
\textbf{\MainTask\ for LLMs.}
\MainTaskAbbr\ methods are typically categorized into two groups: \textit{white-box approaches}, which leverage token-level probabilities and entropy, and \textit{black-box approaches}, which rely solely on final outputs.
Some of these approaches, such as PE~\cite{malinin2021uncertainty}, SE~\cite{kuhn2023semantic}, and Eccentricity~\cite{lin2024generating}, rely on multiple generations of responses, where the core idea is to use the diversity among generated outputs as a proxy for uncertainty. In contrast, another family of approaches operate on a single generation, such as Confidence~\cite{yaldiz2025lars}, which estimates uncertainty by incorporating token probabilities from a single output.
In this work, we focus on white-box probability-based methods and use a single generation to compute the uncertainty score. 

Probability-based \MainTaskAbbr\ methods are based on computing the probability of the generated answer for a given question~\cite{yaldiz2025lars}. 
Formally, given an LLM parameterized by $\theta$ and an input query $q$, the sequence probability is defined as:
\begin{equation}
P(\response \mid \query, \theta)=\prod_{i=1}^N P\left(\response_i \mid \response_{<i}, \query; \theta\right),
\label{eq:sequence_probability}
\end{equation}
where $\response_{<n}$ denotes the tokens generated before the token $\response_n$.
This probability can be directly used as the confidence score of the single most-likely generated answer~\cite{bakman2024mars, yaldiz2025lars, duan2024shifting}, or it can be used to compute the \textit{entropy} of multiple sampled answers using a Monte Carlo approximation~\cite{kuhn2023semantic, malinin2021uncertainty}. 

It has been shown that the sequence probability in Eq.~\eqref{eq:sequence_probability} is biased against longer generations~\cite{malinin2021uncertainty}. To address this issue, length-normalized scoring is introduced as:
$$P_{\text{ln}}(\response \mid \query, \theta) = \prod_{i=1}^N P\left(\response_i \mid \response_{<i}, \query ; \theta\right)^{\frac{1}{N}}.$$
MARS~\cite{bakman2024mars} and TokenSAR~\cite{duan2024shifting} further improve this scoring function by incorporating the semantic contribution of individual tokens. Although their approaches differ in how token importance is modeled, they can be generalized using the following formulation:
$$P_{\text{me}}(\response \mid \query, \theta) = \prod_{i=1}^N P\left(\response_i \mid \response_{<i}, \query ; \theta\right)^{w(\response, \query, N, i)},$$
where $w(\response, \query, N, i)$ denotes the weight assigned to the $i$-th token. These scoring functions assign higher weights to tokens that are central to the answer and are calibrated such that they correlate with the likelihood of error of the final response.
Finally, LARS~\cite{yaldiz2025lars} enhances this calibration by directly learning the scoring function from data.

\vspace{0.5em}
\noindent
\textbf{Learnable Response Scoring (LARS).}
The goal of LARS~\cite{yaldiz2025lars} is to develop a learnable scoring function that captures the semantic contributions of tokens with respect to the query, accounts for biased probabilities, models dependencies between tokens, and incorporates other factors that may not be immediately apparent but are crucial for \MainTaskAbbr.
The key design choice in LARS is feeding probability information into a transformer model, even though this information is a single real value per token. 
Formally, the LARS function is defined as:
\begin{equation}
    P_{\text{LARS}}(r|x, \theta) = \mathcal{S}_{\phi}\big(\query,  \{\langle\response_i, \tilde{p}_i\rangle\}_{i=1}^N; \theta\big),
    \label{eq:lars}
\end{equation}
where $S_{\phi}$ is a trainable scoring function that takes input $\query$ and a set of response tokens $r_i$  and the corresponding probability tokens $\tilde{p}_i$. To train the scoring function $S_{\phi}$, LARS finetunes a pretrained transformer model (specifically RoBERTa-base~\cite{Liu19RoBERTa}). To map real-valued probabilities of answer tokens $r_i$ to high-dimensional vector representations $\tilde{p}_i$ for RoBERTa,   
LARS partitions the probability range $[0,1]$ into $k$ partitions. Given that the transformer model has an input dimension $d$, if $p_i$, the probability of generating $\response_i$, falls within the $r$-th partition, LARS sets its vector positions from $(r - 1) \times \frac{d}{k}$ to $r \times \frac{d}{k}$ to $1$, while the remaining positions are set to $0$. With this encoding strategy, LARS represents distinct probability ranges as orthogonal vectors in a high-dimensional space. The final uncertainty score can be obtained by taking the negative of the score in Eq.~\eqref{eq:lars} for a single generation, or by combining scores from multiple generations, e.g., using entropy-based functions. \citet{yaldiz2025lars} showed that the most-likely generation score outperforms other multi-generation aggregation functions.

\section{Methodology}~\label{sec:methodology}

In this section, we define the problem of \MainTaskAbbr\ for multimodal RAG and present our approach, \textbf{LeMUQ}, a Learnable Multimodal \MainTaskAbbr\ method, building on LARS~\cite{yaldiz2025lars}, which was originally developed for QA tasks with question-only textual inputs.

\subsection{Problem Formulation}
We consider the problem of \MainTaskAbbr\ for responses generated by a VLM in a multimodal RAG setting.
Let $x$ denote the input to the VLM, consisting of: the textual query $\query$, the input image $\image$, and the retrieved textual context $\doc$; i.e., $x = \langle\query, \image, \doc\rangle$. Given this input, a VLM parameterized by $\theta$ generates a response sequence
$\response = (\response_1, \dots, \response_N),$
where $\response_i$ is a token from a vocabulary $\mathcal{V}$, and $N$ is the number of generated tokens.

The uncertainty score is evaluated with respect to the correctness of the response~\cite{uqintroduction}. Let $\response^*$ denote the ground truth response. We compute the correctness of a generated response using exact match (EM), defined as a binary variable $z = \mathbb{I}\{\response = \response^*\},$ where $\mathbb{I}\{\cdot\}$ is the indicator function.

We aim to define an uncertainty scoring function:

$$\mathcal{S}_\phi : (\langle\query, \image, \doc\rangle, \response) \mapsto \mathbb{R},$$ 
parameterized by $\phi$, which assigns higher scores to correct responses than to incorrect ones. That is, for two responses $\response^{(1)}$ and $\response^{(2)}$,
$$ \mathcal{S}_\phi(\langle\query, \image, \doc\rangle, \response^{(1)}) > \mathcal{S}_\phi(\langle\query, \image, \doc\rangle, \response^{(2)}) $$
whenever $\response^{(1)}$ is correct and $\response^{(2)}$ is incorrect. The quality of the uncertainty estimates is then evaluated by measuring how well $\mathcal{S}_\phi$ separates correct from incorrect responses over a dataset; e.g., using metrics such as AUROC~\cite{fawcett2006introduction}.

\subsection{LeMUQ: Learnable Multimodal Uncertainty Quantification}~\label{sec:LeMUQ}
The main idea behind LeMUQ is to model the contribution of different elements of the input to the uncertainty score.
Instead of scoring probabilities based only on the full input, including image and context, we condition on probability estimates under different modality configurations.  In this way, the model can capture uncertainty arising from the visual input, the retrieved context, and their interaction, thereby implicitly learning an uncertainty estimate.


The overall method is illustrated in Figure~\ref{fig:LeMUQ}. First, the full input, consisting of the question, image, and retrieved context, is fed to the VLM to produce a response as a sequence of tokens. For example, given the question ``Where is this bird commonly found?”, the model may generate the answer “Canary Islands.'' Along with the generated tokens, the model also produces token-level probabilities, which are discretized into quantile bins and represented as mapped token probabilities (cf. Sec.~\ref{sec:preliminaries}).
 Second, to disentangle different sources of uncertainty, the same response from the VLM is evaluated under three other input conditions: (1) removing the image, (2) removing the retrieved context, and (3) removing both image and context, using only the question. The token probabilities may shift across these conditions because supporting evidence or noise is removed. 
These alternative probability sequences are likewise converted into mapped token probabilities.
Third, the question and the retrieved context are combined with the generated response and the corresponding mapped token probabilities from all conditions. This joint representation is then passed to a finetuned model, which outputs a single scalar uncertainty score reflecting the likelihood that the generated response is correct, following the Confidence \MainTaskAbbr\ function.

Formally, the scoring function in \proposedmethod\ incorporates four probabilities, where the first is the probability of response $r$ being generated by the VLM $\pi_{\theta}$ given the input $\langle\query, \image, \doc\rangle$:
$$r, p \leftarrow \pi_\theta(\query, \image, \doc ),$$
where $p = (p_1, ..., p_N)$ denotes the probabilities associated with the generation of each token $r_i$ in the response $r = (r_1, ..., r_N)$.
%
Next, we modify the input by removing the context, the image, or both, and compute the probability of generating the same response $r$. Specifically, we construct three variations of the input by:
\begin{enumerate}[leftmargin=*]
\item removing the image: 
$$p^{\text{q,c}} \sim P(r \mid \query, \doc;\theta).$$
\item removing the context: 
$$p^{\text{q, I}} \sim P(r \mid \query, \image;\theta).$$
\item removing both the image and the context: 
$$p^{\text{q}} \sim P(r \mid \query;\theta).$$
\end{enumerate}


The four probability series, $p$, $p^{\text{q,c}}$, $p^{\text{q,I}}$, and $p^{\text{q}}$, are then passed through a probability mapping function to produce the corresponding values $\tilde{p}$, $\tilde{p}^{\text{q,c}}$, $\tilde{p}^{\text{q, I}}$, and $\tilde{p}^{\text{q}}$. We define the \proposedmethod\ score as:
\begin{equation*}
p_{LeMUQ}(r|x, \theta) =
\mathcal{S}_\phi\big(\query,  \doc, \{\langle\response_i, \tilde{p}_i, \tilde{p}^{\text{q,c}}_i, \tilde{p}^{\text{q,I}}_i, \tilde{p}^{\text{q}}_i\rangle\}_{i=1}^N; \theta \big) 
\label{eq:lemuq}
\end{equation*}

This sequence is then fed into a RoBERTa\footnote{\href{https://huggingface.co/docs/transformers/model_doc/roberta}{docs/transformers/model\_doc/roberta}} model, which is finetuned to predict the correctness of the response. The model outputs a single scalar score representing the estimated correctness of the generated response.



\begin{figure}[t]
\centering
\includegraphics[width=0.46\textwidth]{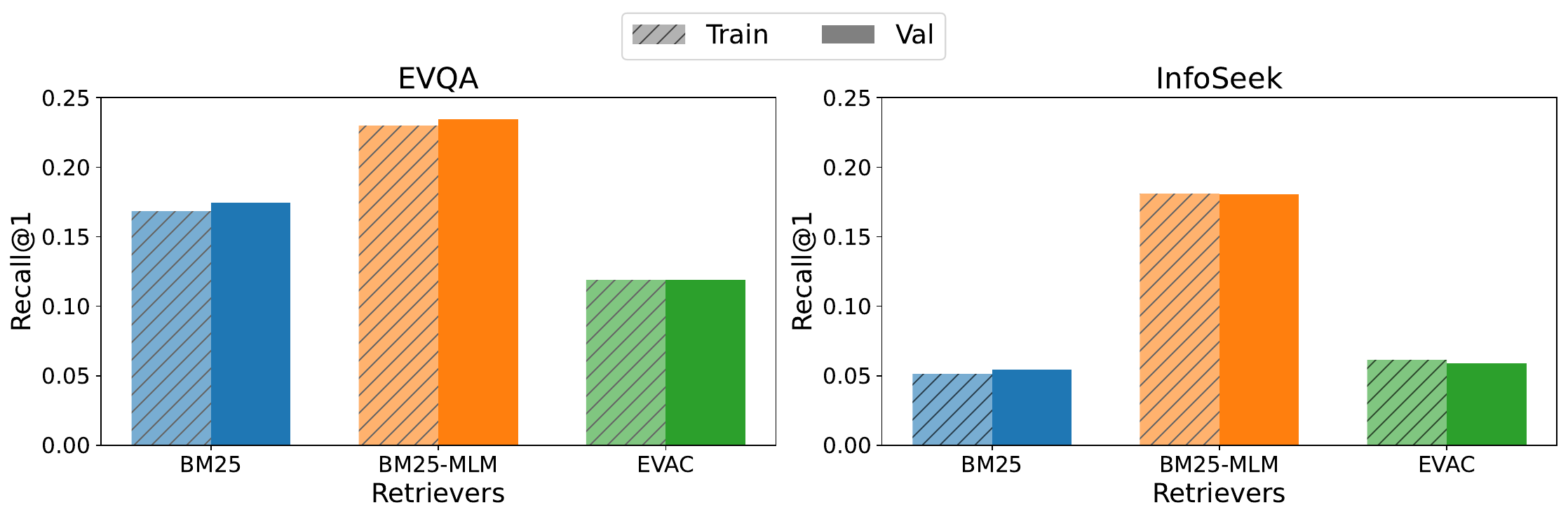}
\caption{Recall@1 performance of retrievers on the train and validation sets.}
\label{fig:recall}
\end{figure}

\renewcommand{\arraystretch}{1.1}
\begin{table*}[t]
\centering
\setlength{\tabcolsep}{2.9pt}
\caption{Within-distribution AUROC performance across EVQA and InfoSeek for LLaVA1.5-7B and Qwen3-VL-4B. BM25, EVAC., and BM25+MLM are used as retrieval models. $\text{LARS}_{Ret}$ uses the finetuned LARS model matched to the corresponding retrieval setting. 
The green values report the gain or loss relative to $\text{LARS}_{Ret}$. Superscripts \textsuperscript{\dag} and \textsuperscript{\ddag} denote statistically significant differences according to the DeLong test ($p<0.05$), compared to $\text{LARS}_{Base}$ and $\text{LARS}_{Ret}$, respectively.}
\label{tab:within_distribution}
\begin{tabular}{l|ll|ll|ll|ll|ll|ll|l}
\hline
\textbf{LLM} & \multicolumn{6}{c|}{\textbf{LLaVA1.5-7B}} & \multicolumn{6}{c|}{\textbf{Qwen3-VL-4B}} & \multirow{3}{*}{\textbf{Avg.}} \\ \cline{0-12}
\textbf{Retriever} & \multicolumn{2}{c}{\textbf{BM25}} & \multicolumn{2}{c}{\textbf{EVAC.}} & \multicolumn{2}{c|}{\textbf{BM25+MLM}} & \multicolumn{2}{c}{\textbf{BM25}} & \multicolumn{2}{c}{\textbf{EVAC.}} & \multicolumn{2}{c|}{\textbf{BM25+MLM}} & \\ \cline{0-12}
\textbf{} & EVQA & InfoSeek & EVQA & InfoSeek & EVQA & InfoSeek & EVQA & InfoSeek & EVQA & InfoSeek & EVQA & InfoSeek & \\ \hline
\rowcolor{gray!15} Accuracy & 0.128 & 0.118 & 0.138 & 0.102 & 0.163 & 0.147 & 0.133 & 0.142 & 0.135 & 0.127 & 0.169 & 0.140 & 0.137 \\
PE & 0.750 & 0.756 & 0.803 & 0.775 & 0.719 & 0.729 & 0.721 & 0.770 & 0.753 & 0.813 & 0.692 & 0.749 & 0.752 \\
P(True) & 0.639 & 0.645 & 0.677 & 0.645 & 0.641 & 0.614 & 0.604 & 0.734 & 0.596 & 0.714 & 0.584 & 0.740 & 0.653 \\
Ecc. & 0.732 & 0.745 & 0.752 & 0.755 & 0.699 & 0.717 & 0.664 & 0.667 & 0.728 & 0.705 & 0.634 & 0.629 & 0.702 \\
Img. Per. & 0.415 & 0.414 & 0.426 & 0.409 & 0.383 & 0.397 & 0.347 & 0.294 & 0.348 & 0.321 & 0.389 & 0.291 & 0.369 \\
$\text{LARS}_{Base}$ & 0.689 & 0.762 & 0.709 & 0.752 & 0.687 & 0.752 & 0.673 & 0.765 & 0.767 & 0.825 & 0.607 & 0.771 & 0.730 \\
$\text{LARS}_{Ret}$ & 0.847 & 0.829 & 0.848 & 0.823 & 0.846 & 0.791 & 0.840 & 0.858 & 0.877 & 0.890 & 0.816 & 0.845 & 0.843 \\
\textbf{\proposedmethod} & \textbf{0.855}\textsuperscript{\dag} & \textbf{0.833}\textsuperscript{\dag} & \textbf{0.861}\textsuperscript{\dag} & \textbf{0.854}\textsuperscript{\dag}\textsuperscript{\ddag} & \textbf{0.871}\textsuperscript{\dag}\textsuperscript{\ddag} & \textbf{0.853}\textsuperscript{\dag}\textsuperscript{\ddag} & \textbf{0.923}\textsuperscript{\dag}\textsuperscript{\ddag} & \textbf{0.888}\textsuperscript{\dag}\textsuperscript{\ddag} & \textbf{0.912}\textsuperscript{\dag}\textsuperscript{\ddag} & \textbf{0.913}\textsuperscript{\dag}\textsuperscript{\ddag} & \textbf{0.902}\textsuperscript{\dag}\textsuperscript{\ddag} & \textbf{0.898}\textsuperscript{\dag}\textsuperscript{\ddag} & \textbf{0.880} \\
 & \scriptsize \textcolor{ForestGreen}{(+0.008)} & \scriptsize \textcolor{ForestGreen}{(+0.004)} & \scriptsize \textcolor{ForestGreen}{(+0.012)} & \scriptsize \textcolor{ForestGreen}{(+0.031)} & \scriptsize \textcolor{ForestGreen}{(+0.025)} & \scriptsize \textcolor{ForestGreen}{(+0.063)} & \scriptsize \textcolor{ForestGreen}{(+0.083)} & \scriptsize \textcolor{ForestGreen}{(+0.030)} & \scriptsize \textcolor{ForestGreen}{(+0.034)} & \scriptsize \textcolor{ForestGreen}{(+0.022)} & \scriptsize \textcolor{ForestGreen}{(+0.086)} & \scriptsize \textcolor{ForestGreen}{(+0.053)} & \scriptsize \textcolor{ForestGreen}{(+0.038)} \\
\hline
\end{tabular}
\miniskip
\end{table*}

\vspace{0.5em}
\noindent
\textbf{Training Strategy.} Following LARS, let $\mathcal{S}_\phi$ denote the scoring function, which takes four inputs: the input query $\query$, the retrieved context $\doc$ , the generated sequence $\response = (\response_1, \dots, \response_N)$, and the corresponding mapped probability vector, 
$\tilde{p} = (\tilde{p}_1, \tilde{p}^{\text{q,c}}_1, \tilde{p}^{\text{q,I}}_1, \tilde{p}^{\text{q}}_1, \dots,\tilde{p}_N, \tilde{p}^{\text{q,c}}_N,$ $ \tilde{p}^{\text{q,I}}_N, \tilde{p}^{\text{q}}_N,),$
where $\tilde{p}_{i},\tilde{p}^{\text{q,c}}_i, \tilde{p}^{\text{q,I}}_i, \tilde{p}^{\text{q}}_i$ represent the mapped probability tokens of token $\response_{i}$. The function $\mathcal{S}_\phi$ outputs a real-valued score~$o$. We make $\mathcal{S}_\phi$ directly learnable using supervised data.
To this end, we construct a calibration set to train the scoring function $\mathcal{S}_\phi$, parameterized by $\phi$. 
This calibration set consists of 5-tuples: the input query $\query$, the retrieved context $\doc$, the generated sequence $\response$, the mapped probability vector $\tilde{p}$, and a binary ground truth label $g$. The label $g$ indicates whether $\response$ is a correct response to $\query$. To optimize the parameters of $\mathcal{S}_\phi$, we employ the binary cross-entropy loss. 
\section{Experimental Setup}~\label{sec:experimental_setup}

\renewcommand{\arraystretch}{1.2}
\begin{table*}[t]
\centering
\setlength{\tabcolsep}{6pt}
\caption{Generalizability of \proposedmethod\ across different retrieval models. The UQ model is trained using one retrieval model and tested with another retrieval model. Superscripts \textsuperscript{\dag} and \textsuperscript{\ddag} denote statistically significant differences according to the DeLong test ($p<0.05$), compared to $\text{LARS}_{Base}$ and the corresponding LARS model in the same row block, respectively.}
\label{tab:ood_retriever}
\begin{tabular}{l|lllll|lllll
}
\hline
\textbf{VLM} & \multicolumn{5}{c|}{\textbf{EVQA}} & \multicolumn{5}{c}{\textbf{InfoSeek}} \\ \hline
\textbf{} &
\textbf{$\text{Doc}^{\times}$} & \textbf{BM25} & \textbf{EVAC.} & \shortstack[c]{\textbf{BM25+}\\\textbf{MLM}} & \textbf{$\text{Doc}^{+}$} & 
\textbf{$\text{Doc}^{\times}$} & \textbf{BM25} & \textbf{EVAC.} & \shortstack[c]{\textbf{BM25+}\\\textbf{MLM}} & \textbf{$\text{Doc}^{+}$} \\ \hline
\multicolumn{11}{l}{\textit{LLaVA1.5-7B}} \\
\hline
LARS$_{Base}$ & 0.745 & 0.689 & 0.709 & 0.687 & 0.674 & 0.773 & 0.762 & 0.752 & 0.752 & 0.690 \\
\rowcolor{gray!15} LARS$_{BM25}$ & 0.847 & --- & 0.804 & 0.835 & 0.701 & 0.909 & --- & 0.845 & 0.797 & 0.779 \\
\rowcolor{gray!15} \proposedmethod$_{BM25}$ & 0.823\textsuperscript{\dag} & --- & 0.828\textsuperscript{\dag}\textsuperscript{\ddag} & 0.841\textsuperscript{\dag} & 0.739\textsuperscript{\dag}\textsuperscript{\ddag} & 0.824\textsuperscript{\ddag} & --- & 0.819\textsuperscript{\dag} & 0.810\textsuperscript{\dag} & 0.795\textsuperscript{\dag} \\
LARS$_{EVAC}$ & 0.867 & 0.807 & --- & 0.812 & 0.813 & 0.896 & 0.825 & --- & 0.788 & 0.783 \\
\proposedmethod$_{EVAC}$ & 0.832\textsuperscript{\dag}\textsuperscript{\ddag} & 0.799\textsuperscript{\dag} & --- & 0.791\textsuperscript{\dag} & 0.796\textsuperscript{\dag}\textsuperscript{\ddag} & 0.866\textsuperscript{\dag} & 0.828\textsuperscript{\dag} & --- & 0.814\textsuperscript{\dag}\textsuperscript{\ddag} & 0.800\textsuperscript{\dag} \\
\rowcolor{gray!15}\shortstack[c]{LARS$_{\mathrm{BM25{+}MLM}}$} & 0.827 & 0.844 & 0.799 & --- & 0.716 & 0.917 & 0.822 & 0.849 & --- & 0.762 \\
\rowcolor{gray!15}\shortstack[c]{\proposedmethod$_{\mathrm{BM25{+}MLM}}$} & 0.821\textsuperscript{\dag} & 0.860\textsuperscript{\dag} & 0.840\textsuperscript{\dag}\textsuperscript{\ddag} & --- & 0.755\textsuperscript{\dag}\textsuperscript{\ddag} & 0.802\textsuperscript{\ddag} & 0.823\textsuperscript{\dag} & 0.824\textsuperscript{\dag} & --- & 0.785\textsuperscript{\dag}\textsuperscript{\ddag} \\

\hline
\multicolumn{11}{l}{\textit{Qwen3-VL-4B}} \\
\hline
LARS$_{Base}$ & 0.709 & 0.673 & 0.767 & 0.607 & 0.629 & 0.751 & 0.765 & 0.825 & 0.771 & 0.772 \\

\rowcolor{gray!15}LARS$_{BM25}$ & 0.869 & --- & 0.893 & 0.811 & 0.841 & 0.916 & --- & 0.893 & 0.845 & 0.854 \\
\rowcolor{gray!15}\proposedmethod$_{BM25}$ & 0.898\textsuperscript{\dag}\textsuperscript{\ddag} & --- & 0.889\textsuperscript{\dag} & 0.904\textsuperscript{\dag}\textsuperscript{\ddag} & 0.881\textsuperscript{\dag}\textsuperscript{\ddag} & 0.879\textsuperscript{\dag}\textsuperscript{\ddag} & --- & 0.891\textsuperscript{\dag} & 0.896\textsuperscript{\dag}\textsuperscript{\ddag} & 0.871\textsuperscript{\dag} \\
LARS$_{EVAC}$ & 0.879 & 0.810 & --- & 0.777 & 0.815 & 0.913 & 0.854 & --- & 0.843 & 0.852 \\
\proposedmethod$_{EVAC}$ & 0.857\textsuperscript{\dag}\textsuperscript{\ddag} & 0.878\textsuperscript{\dag}\textsuperscript{\ddag} & --- & 0.857\textsuperscript{\dag}\textsuperscript{\ddag} & 0.903\textsuperscript{\dag}\textsuperscript{\ddag} & 0.894\textsuperscript{\dag}\textsuperscript{\ddag} & 0.894\textsuperscript{\dag}\textsuperscript{\ddag} & --- & 0.906\textsuperscript{\dag}\textsuperscript{\ddag} & 0.888\textsuperscript{\dag}\textsuperscript{\ddag} \\
\rowcolor{gray!15}\shortstack[c]{LARS$_{\mathrm{BM25{+}MLM}}$} & 0.882 & 0.846 & 0.897 & --- & 0.856 & 0.914 & 0.854 & 0.887 & --- & 0.862 \\
\rowcolor{gray!15}\shortstack[c]{\proposedmethod$_{\mathrm{BM25{+}MLM}}$} & 0.821\textsuperscript{\dag}\textsuperscript{\ddag} & 0.913\textsuperscript{\dag}\textsuperscript{\ddag} & 0.894\textsuperscript{\dag} & --- & 0.898\textsuperscript{\dag}\textsuperscript{\ddag} & 0.873\textsuperscript{\dag}\textsuperscript{\ddag} & 0.882\textsuperscript{\dag}\textsuperscript{\ddag} & 0.899\textsuperscript{\dag} & --- & 0.874\textsuperscript{\dag} \\
\hline
\end{tabular}
\end{table*}

\noindent
\textbf{Datasets.}
As multimodal RAG requires $\langle$question, image, answer$\rangle$ triplets that also rely on external knowledge, we evaluate \proposedmethod\ on two vision-text datasets: Encyclopedic VQA (EVQA)~\cite{mensink2023encyclopedic} and InfoSeek~\cite{chen2023infoseek}. Both datasets come with a controlled knowledge base generated from Wikipedia articles containing ground truth documents for each question.

EVQA focuses on encyclopedic knowledge about entities such as landmarks and species, and includes various question types. Following~\cite{yan2024echosight}, we use only the automatically generated questions and ensure each question–entity pair appears exclusively in either train or test set. Due to limitations in the original splits, we subsample the training data to create 15,000 training and 2,000 test examples, ensuring no overlap between the train and test splits.

InfoSeek similarly targets questions requiring external knowledge. Since its required articles are covered by the EVQA knowledge base, we use the same source for retrieval. In contrast to EVQA, only the ground truth Wikipedia article is given for each question. To identify relevant sections, we apply a MiniLM model trained on MS MARCO\footnote{\href{https://huggingface.co/cross-encoder/ms-marco-MiniLM-L-6-v2}{cross-encoder/ms-marco-MiniLM-L-6-v2}} that ranks sections using the question and answer. 
From the training split, we sample 15,000 instances for training, ensuring that all unique question–entity pairs are preserved. For evaluation, we sample 2,000 instances from the validation split.

\vspace{0.5em}
\noindent
\textbf{Retrieval Models.}
Following~\cite{soudani2025uncertainty}, we use a BM25 retriever~\cite{BM2509Robertson} and a two-stage reranking pipeline which combines BM25 with the MS MARCO cross-encoder reranker, which is applied to the top 1000 retrieved documents (BM25+MiniLM). In addition to text-only retrieval methods, we also perform image-to-image retrieval~\cite{cocchi2025augmenting} by encoding query images with EVA-CLIP-8B\footnote{\href{https://huggingface.co/BAAI/EVA-CLIP-8B}{BAAI/EVA-CLIP-8B}} and matching them to Wikipedia image embeddings to retrieve relevant articles. We then use Contriever\footnote{\href{https://huggingface.co/facebook/contriever}{facebook/contriever}} to select the most relevant section given the question (EVAC.). We also consider an oracle retrieval setting $Doc^{+}$ (gold document), and no retrieval setting  $Doc^{\times}$ (no document).

On InfoSeek, standard BM25-based methods yield low recall. To address this, we adopt a query rewriting strategy inspired by~\cite{ma2023query}, where GPT-5-mini\footnote{\href{https://developers.openai.com/api/docs/models/gpt-5-mini}{https://developers.openai.com/api/docs/models/gpt-5-mini}} generates a query combining the visual entity and question intent. This rewritten query is used for retrieval, and for BM25+MiniLM, we perform the second-stage retrieval using the identified entity in the image instead of the abstract entity in the original question. These modifications improve retrieval quality and ensure comparability across datasets.

\vspace{0.5em}
\noindent
\textbf{VLMs.}
To generate responses, we use two VLMs: LLaVA1.5-7B\footnote{\href{https://huggingface.co/llava-hf/llava-1.5-7b-hf}{llava-hf/llava-1.5-7b-hf}} and Qwen3-VL-4B.\footnote{\href{https://huggingface.co/Qwen/Qwen3-VL-4B-Instruct}{Qwen/Qwen3-VL-4B-Instruct}}
For response generation, we set the \texttt{do\_sample} parameter to \texttt{False}. For sampling responses for UQ methods, we set the number of samples to 10, enable \texttt{do\_sample}, and set both temperature and top\_p to 1.0.
We provide the top-1 retrieved section as context to the VLM following the simplified setup in~\cite{cocchi2025augmenting}. Following prior work in RAG~\cite{Soudani24FTvsRAG, mallen23popqa}, which limits the effective input by restricting retrieved content~\cite{rag}, we constrain the input length to fit within hardware limits. In particular, we truncate the context to a maximum of approximately 500 tokens.


\vspace{0.5em}
\noindent
\textbf{Baselines.}
We use the following baseline \MainTaskAbbr\ methods as discussed in Sections \ref{sec:related_work} and \ref{sec:preliminaries}: Ecc~\cite{lin2024generating}, PE~\cite{malinin2021uncertainty}, P(True)~\cite{kadavath2022language}, and the pretrained LARS (\LARSBase). To account for visual uncertainty, we include the image perturbation (Img. Per.) method~\cite{avestimehr2025detecting}, described in Section~\ref{sec:related_work}. We instantiate this approach using a black image as the perturbed version and compute the distance on the top-1 token distribution, as this was shown to achieve the best performance~\cite{avestimehr2025detecting}. In addition to the baseline methods, we finetune LARS and our proposed LeMUQ. For response scoring, we use a single-generation setup (denoted as Confidence in~\cite{yaldiz2025lars}), as it achieves the best performance among probability-based UQ methods in~\cite{yaldiz2025lars}. 


\vspace{0.5em}
\noindent
\textbf{Evaluation Metrics.}
To evaluate the responses, we use exact matching as in~\cite{soudani2025uncertainty}. Similarly, we use AUROC as an evaluation metric for the UQ methods, as it is invariant to the distribution of correct and incorrect answers.

\vspace{0.5em}
\noindent
\textbf{Finetuning Setup.}
To finetune LARS and \proposedmethod, we follow the training pipeline of~\cite{yaldiz2025lars}. For each $\langle$question, image, answer$\rangle$ triplet, we retrieve context using one of the previously defined retrievers and generate up to five responses with the VLM, ensuring that the most likely response is included. Each generated response is labeled using exact matching against the ground-truth answer. As the scoring model, we use a RoBERTa-base model initialized with the finetuned LARS weights from~\cite{yaldiz2025lars} for both LARS and LeMUQ. We train the models to predict the correctness of the generated response using binary cross-entropy with a learning rate of $5\times10^{-6}$ for three epochs. 10\% of the training data is used for validation. Training was conducted on RTX A6000, H100, and H200 GPUs, with training times ranging from 10 to 27 hours for LeMUQ and 3–8 hours for LARS, depending on the retriever, VLM, and hardware used.

\section{Results}

We present a set of experiments that address the following research questions:
\textbf{RQ1}: How does our proposed method perform compared to baseline UQ methods when evaluated within-distribution? 
\textbf{RQ2}: How well does \proposedmethod\ generalize to out-of-distribution data? 
\textbf{RQ3}: What is the impact of each probability component in \proposedmethod\ on its overall performance?

\subsection{\proposedmethod\ Performance}~\label{sec:LeMUQ_performance}
\textbf{RQ1} evaluates the performance of \proposedmethod\ compared to other \MainTaskAbbr\ methods in a within-distribution setting. 
Figure~\ref{fig:recall} presents the performance of the retrievers on both the training and validation sets. We report Recall@1, as only the top-ranked document is passed to the generator model. The BM25+MLM model achieves the best performance on both datasets.
For the InfoSeek dataset, BM25 and EVAC perform similarly; however, BM25 outperforms EVAC on the EVQA dataset. The relatively lower performance of EVAC can be attributed to its two-stage retrieval pipeline, which first operates at the article level and then at the section level. As a result, the overall performance is constrained by the effectiveness of the initial article-level retrieval step.
In summary, we conduct our experiments using different types of retrievers with varying levels of performance.

Table \ref{tab:within_distribution} presents the results on two datasets and two VLMs. By \textit{within-distribution}, we mean that the model is trained and evaluated using the same retrieval model; in other words, the documents used for training and evaluation are obtained from the same retriever.
Among the baselines, PE, P(True), Ecc., and \LARSBase\ perform reasonably well and are also robust to changes. Img. Per. fails to predict uncertainty properly, consistently returning values below 0.5. This indicates that Img. Per. is not generalizable to multimodal RAG settings. By finetuning the standard LARS model for a specific dataset, VLM, and retriever, \LARSRet\ outperforms baseline methods.

Finally, our proposed method, \proposedmethod, outperforms all baselines across all settings by a significant margin. Overall, it increases the AUROC by $3.8\%$ on average compared to a finetuned LARS scoring function. The improvement achieved by \proposedmethod\ is higher in the Qwen3 model, with an average improvement of $5.1\%$, compared to the improvement observed in the LLaVA1.5 model, which is on average $2.4\%$. When comparing across datasets, our method consistently outperforms the baselines across all types of retrievers, including sparse, reranking, and image-based retrievers. However, the performance of the UQ methods is higher for the EVAC retriever on average across both datasets and models.
These findings suggest that incorporating the probability of VLM generation under different input configurations, and training a model to capture these signals, improves uncertainty estimation in multimodal RAG systems.

\renewcommand{\arraystretch}{1.1}
\begin{table}[t]
\centering
\setlength{\tabcolsep}{1.4pt}
\caption{Generalizability of LeMUQ across datasets. The UQ model is trained on one dataset and tested on another dataset. Superscript \textsuperscript{\ddag} denotes a statistically significant difference according to the DeLong test ($p<0.05$) compared to $\text{LARS}_{Ret.}$ in the same column.}
\label{tab:ood_dataset}
\begin{tabular}{>{\raggedright\arraybackslash}m{1.2cm}|>{\centering\arraybackslash}m{1.1cm}>{\centering\arraybackslash}m{1.1cm}>{\centering\arraybackslash}m{1.1cm}|>{\centering\arraybackslash}m{1.1cm}>{\centering\arraybackslash}m{1.1cm}>{\centering\arraybackslash}m{1.1cm}}
\hline
\textbf{Setup} & \multicolumn{3}{c|}{\textbf{EVQA $\rightarrow$ InfoSeek}} & \multicolumn{3}{c}{\textbf{InfoSeek $\rightarrow$ EVQA}} \\ \hline
\textbf{Ret.} & \textbf{BM25} & \textbf{EVAC.} & \shortstack[c]{\textbf{BM25+}\\\textbf{MLM}} & \textbf{BM25} & \textbf{EVAC.} & \shortstack[c]{\textbf{BM25+}\\\textbf{MLM}} \\ \hline
\multicolumn{7}{c}{\textit{LLaVA1.5-7B}} \\ \hline
$\text{LARS}_{Ret.}$ & 0.632 & 0.803 & 0.621 & 0.618 & 0.700 & 0.597 \\
\shortstack[c]{\textbf{$\text{LeMUQ}$}} & 0.768\textsuperscript{\ddag} & 0.812 & 0.721\textsuperscript{\ddag} & 0.646 & 0.720 & 0.657\textsuperscript{\ddag} \\
\hline
\multicolumn{7}{c}{\textit{Qwen3-VL-4B}} \\\hline
$\text{LARS}_{Ret.}$ & 0.813 & 0.845 & 0.805 & 0.718 & 0.754 & 0.601 \\
\shortstack[c]{\textbf{$\text{LeMUQ}$}} & 0.820 & 0.839 & 0.852\textsuperscript{\ddag} & 0.778\textsuperscript{\ddag} & 0.830\textsuperscript{\ddag} & 0.755\textsuperscript{\ddag} \\

\hline
\end{tabular}
\end{table}

\subsection{Out-of-Domain Generalizability}~\label{sec:ood_performance}
\textbf{RQ2} evaluates the out-of-distribution (OOD) generalizability of \proposedmethod\ across three settings: retriever, dataset, and VLM. The main objective is to assess whether our proposed scoring function generalizes effectively to these different settings. 


\smallskip\noindent\textbf{Retriever Generalizability.}
Table~\ref{tab:ood_retriever} presents the results of LeMUQ when the model is trained on contexts retrieved by one retriever and evaluated on contexts obtained from different retrievers. The rows correspond to the retrievers used during training, while the columns represent the retrievers used at evaluation time. In addition to evaluating with BM25, BM25+MLM, and EVAC, we also consider two special settings: one without any context in the input ($Doc^\times$), and another where the input contains gold context ($Doc^+$).

For LLaVA1.5 on the EVQA dataset, when training is performed using BM25 or BM25+MLM, \proposedmethod\ consistently outperforms both \LARSBase\ and \LARSRet\ across BM25, EVAC, BM25+MLM, and $Doc^+$. However, when training is conducted using EVAC, \LARSRet\ surpasses \proposedmethod. Additionally, \LARSRet\ performs better than \proposedmethod\ in the $Doc^\times$ setting. This suggests that, under our training approach, \proposedmethod\ is less effective when no context is provided, although it still significantly outperforms \LARSBase. 
For InfoSeek, \LARSRet\ only outperforms \proposedmethod\ in the $Doc^\times$ setting; in all other configurations, \proposedmethod\ shows a significant advantage over the baselines.

For Qwen3-VL, the results are largely consistent with our observations for LLaVA1.5. However, on the EVQA dataset, \proposedmethod\ outperforms \LARSRet\ in the $Doc^\times$ setting when trained with BM25.
Across both datasets, the generalization performance is significantly higher compared to what we observed with LLaVA1.5. For example, when training on EVQA and evaluating on BM25+MLM, the improvement is 0.093.
Overall, these findings indicate that although our scoring function experiences a drop in performance in the absence of context, it still generalizes well across different retrievers compared to the baseline. This suggests that the scoring function is relatively insensitive to the choice of retriever.



\renewcommand{\arraystretch}{1.1}
\begin{table}[t]
\centering
\setlength{\tabcolsep}{1.4pt}
\caption{Generalizability of LeMUQ across VLMs. The UQ model is trained on one VLM and tested on another VLM. Superscript \textsuperscript{\ddag} denotes a statistically significant difference according to the DeLong test ($p<0.05$) compared to $\text{LARS}_{Ret.}$ in the same column.}
\label{tab:ood_vlm}
\begin{tabular}{>{\raggedright\arraybackslash}m{1.2cm}|>{\centering\arraybackslash}m{1.1cm}>{\centering\arraybackslash}m{1.1cm}>{\centering\arraybackslash}m{1.1cm}|>{\centering\arraybackslash}m{1.1cm}>{\centering\arraybackslash}m{1.1cm}>{\centering\arraybackslash}m{1.1cm}}
\hline
\textbf{Setup} & \multicolumn{3}{c|}{\textbf{LLaVA $\rightarrow$ Qwen3}} & \multicolumn{3}{c}{\textbf{Qwen3 $\rightarrow$ LLaVA}} \\ \hline
\textbf{Ret.} & \textbf{BM25} & \textbf{EVAC.} & \shortstack[c]{\textbf{BM25+}\\\textbf{MLM}} & \textbf{BM25} & \textbf{EVAC.} & \shortstack[c]{\textbf{BM25+}\\\textbf{MLM}} \\ \hline
\multicolumn{7}{c}{\textit{EVQA}} \\ \hline
$\text{LARS}_{Ret.}$ & 0.812 & 0.786 & 0.790 & 0.761 & 0.776 & 0.729 \\
\shortstack[c]{\textbf{$\text{LeMUQ}$}} & 0.842\textsuperscript{\ddag} & 0.824\textsuperscript{\ddag} & 0.801 & 0.716\textsuperscript{\ddag} & 0.695\textsuperscript{\ddag} & 0.718 \\
\hline
\multicolumn{7}{c}{\textit{InfoSeek}} \\\hline
$\text{LARS}_{Ret.}$ & 0.757 & 0.786 & 0.748 & 0.805 & 0.808 & 0.775 \\
\shortstack[c]{\textbf{$\text{LeMUQ}$}} & 0.811\textsuperscript{\ddag} & 0.838\textsuperscript{\ddag} & 0.852\textsuperscript{\ddag} & 0.678\textsuperscript{\ddag} & 0.715\textsuperscript{\ddag} & 0.669\textsuperscript{\ddag} \\

\hline
\end{tabular}
\miniskip
\end{table}

\renewcommand{\arraystretch}{1.15}
\begin{table*}[t]
\centering
\setlength{\tabcolsep}{2.9pt}
\caption{Ablation study AUROC performance across EVQA and InfoSeek for LLaVA1.5-7B and Qwen3-VL-4B. BM25, EVAC., and BM25+MLM are used as retrieval models. LeMUQ denotes the full method, and the remaining rows remove one component at a time. Superscripts \textsuperscript{\dag} and \textsuperscript{\ddag} denote statistically significant differences according to the DeLong test ($p<0.05$), compared to $\text{LARS}_{Ret}$ and LeMUQ, respectively.}
\label{tab:lemuq_ablation}
\begin{tabular}{l|ll|ll|ll|ll|ll|ll|l}
\hline
\textbf{VLM} & \multicolumn{6}{c|}{\textbf{LLaVA1.5-7B}} & \multicolumn{6}{c|}{\textbf{Qwen3-VL-4B}} & \multirow{3}{*}{\textbf{Avg.}} \\ \cline{1-13}
\textbf{Retriever} & \multicolumn{2}{c}{\textbf{BM25}} & \multicolumn{2}{c}{\textbf{EVAC.}} & \multicolumn{2}{c|}{\textbf{BM25+MLM}} & \multicolumn{2}{c}{\textbf{BM25}} & \multicolumn{2}{c}{\textbf{EVAC.}} & \multicolumn{2}{c|}{\textbf{BM25+MLM}} & \\ \cline{1-13}
\textbf{} & EVQA & InfoSeek & EVQA & InfoSeek & EVQA & InfoSeek & EVQA & InfoSeek & EVQA & InfoSeek & EVQA & InfoSeek & \\ \hline
\rowcolor{gray!15} Accuracy & 0.128 & 0.118 & 0.138 & 0.102 & 0.163 & 0.147 & 0.136 & 0.142 & 0.132 & 0.128 & 0.171 & 0.140 & 0.137 \\
$\text{LARS}_{Base}$ & 0.689 & 0.762 & 0.709 & 0.752 & 0.687 & 0.752 & 0.673 & 0.767 & 0.769 & 0.815 & 0.610 & 0.773 & 0.730 \\
$\text{LARS}_{Ret}$ & 0.847 & 0.829 & 0.848 & 0.823 & 0.846 & 0.791 & 0.832 & 0.861 & 0.882 & 0.889 & 0.816 & 0.843 & 0.842 \\
\textbf{LeMUQ} & 0.855 & 0.832 & \textbf{0.861} & \textbf{0.854}\textsuperscript{\dag} & \textbf{0.871}\textsuperscript{\dag} & 0.853\textsuperscript{\dag} & 0.920\textsuperscript{\dag} & 0.889\textsuperscript{\dag} & 0.917\textsuperscript{\dag} & 0.909\textsuperscript{\dag} & 0.902\textsuperscript{\dag} & 0.901\textsuperscript{\dag} & \textbf{0.880} \\
LeMUQ $-\tilde{p}$ & 0.841 & 0.826 & 0.852 & 0.825\textsuperscript{\ddag} & 0.859 & 0.844\textsuperscript{\dag} & 0.921\textsuperscript{\dag} & 0.903\textsuperscript{\dag}\textsuperscript{\ddag} & 0.910\textsuperscript{\dag} & 0.906\textsuperscript{\dag} & 0.894\textsuperscript{\dag} & \textbf{0.909}\textsuperscript{\dag} & 0.874 \\
LeMUQ $-\tilde{p}^{q,I}$ & 0.856 & 0.845 & 0.851 & 0.829\textsuperscript{\ddag} & 0.840\textsuperscript{\ddag} & 0.829\textsuperscript{\dag}\textsuperscript{\ddag} & \textbf{0.924}\textsuperscript{\dag} & 0.891\textsuperscript{\dag} & 0.912\textsuperscript{\dag} & 0.906\textsuperscript{\dag} & \textbf{0.907}\textsuperscript{\dag} & 0.893\textsuperscript{\dag} & 0.873 \\
LeMUQ $-\tilde{p}^{q,c}$ & 0.852 & 0.818 & 0.852 & 0.849\textsuperscript{\dag} & 0.856 & \textbf{0.862}\textsuperscript{\dag} & 0.923\textsuperscript{\dag} & \textbf{0.906}\textsuperscript{\dag}\textsuperscript{\ddag} & \textbf{0.919}\textsuperscript{\dag} & 0.899 & 0.895\textsuperscript{\dag} & 0.905\textsuperscript{\dag} & 0.878 \\
LeMUQ $-\tilde{p}^{q}$ & \textbf{0.859} & \textbf{0.849} & 0.851 & 0.848\textsuperscript{\dag} & 0.864 & 0.851\textsuperscript{\dag} & 0.906\textsuperscript{\dag}\textsuperscript{\ddag} & 0.902\textsuperscript{\dag}\textsuperscript{\ddag} & 0.916\textsuperscript{\dag} & \textbf{0.911}\textsuperscript{\dag} & 0.886\textsuperscript{\dag}\textsuperscript{\ddag} & 0.905\textsuperscript{\dag} & 0.879 \\
\hline
\end{tabular}
\miniskip
\end{table*}

\smallskip\noindent\textbf{Dataset Generalizability.}
Table~\ref{tab:ood_dataset} presents the results of dataset generalization. In this setup, the model is trained on one dataset and evaluated on a different dataset. In the table, the dataset on the left side of the arrow indicates the training dataset, while the one on the right side of the arrow represents the evaluation dataset.
To isolate the effect of the dataset, we keep both the retriever and the VLM the same during training and evaluation.

On LLaVA1.5-7B, \proposedmethod\ consistently outperforms the baseline across all setups. Moreover, when using BM25+MLM, \proposedmethod\ significantly outperforms \LARSRet.
On Qwen3-VL-4B, the baseline performs better in only one out of six setups, which demonstrates the strong generalization ability of our model across datasets.
Based on these observations, we conclude that the proposed scoring function performs well even when evaluated on datasets different from those used during training, including those with different distributions. This suggests that the scoring function learns to effectively utilize probability information itself, rather than relying on dataset-specific distributions.

\smallskip\noindent\textbf{VLM Generalizability.}
Table~\ref{tab:ood_vlm} demonstrates the generalization ability of \MainTaskAbbr\ methods across different VLMs. Similar to other generalization experiments, we fix both the dataset and the retriever during training and evaluation, and assess cross-model generalization from Qwen to LLaVA and vice versa.

From LLaVA to Qwen3, across both datasets and all retrievers, \proposedmethod\ consistently outperforms the baseline. This indicates that a model trained on LLaVA generalizes well to Qwen.
In contrast, from Qwen3 to LLaVA, the baseline performs better than our proposed method. This suggests that training on Qwen leads the scoring function to learn the probability distribution specific to the Qwen model, which does not transfer effectively to LLaVA.
Overall, the results show that the generalization capability of \proposedmethod\ depends on the choice of training and evaluation backbone models.

\subsection{Impact of Probability Components}~\label{sec:probability_impact}
\proposedmethod\ consists of four probability components: $\tilde{p}'$, $\tilde{p}^{\text{q,c}}$, $\tilde{p}^{\text{q,I}}$, and $\tilde{p}^{\text{q}}$, as introduced in Section~\ref{sec:LeMUQ}. \textbf{RQ3} examines the contribution of each component to overall performance by performing an ablation study, where one component is removed at a time, followed by finetuning and evaluation using the remaining three components.

Table~\ref{tab:lemuq_ablation} summarizes the findings of the study.
For LLaVA1.5-7B combined with BM25, the highest performance is achieved when $p^{\text{q}}$ is excluded, followed by the configuration without $p^{\text{q,I}}$ across both datasets.
In contrast, when using the EVAC retriever, \proposedmethod\ with all components delivers the strongest results. In particular, on the InfoSeek dataset, removing $\tilde{p}'$ or $p^{\text{q,I}}$ leads to a substantial decline in performance.
For the BM25+MLM setting, \proposedmethod\ again yields the best overall results on average; however, omitting $p^{\text{q,I}}$ results in a pronounced degradation.

On Qwen3-VL-4B with BM25, performance improves slightly when excluding $p^{\text{q,I}}$, $p^{\text{q,c}}$, and $\tilde{p}'$, whereas removing $p^{\text{q}}$ leads to a substantial decline. For EVQA, \proposedmethod\ achieves the second-best results. Under the BM25+MLM setting, the highest scores are obtained by omitting $p^{\text{q,I}}$ for EVQA and $\tilde{p}'$ for InfoSeek. Across most configurations, variants of \proposedmethod\ consistently outperform $\text{LARS}_{Ret}$, highlighting the effectiveness and necessity of the proposed probability components.

Overall, \proposedmethod\ delivers the strongest average performance, suggesting that each component contributes positively to the final outcome. Notably, removing $p^{\text{q,I}}$ and $\tilde{p}'$ results in larger performance drops compared to excluding $p^{\text{q,c}}$ or $p^{\text{q}}$, indicating that these two components play a more critical role in achieving high performance.

\section{Conclusions and Future Work}~\label{sec:conclusions}
This paper investigates the problem of uncertainty quantification for multimodal RAG for the first time, to the best of our knowledge. It  introduces a novel method for \MainTaskAbbr\, called \proposedmethod, designed specifically for the multimodal RAG setting. \proposedmethod\ is a learnable scoring approach that analyzes token probabilities under various input modifications, such as removing modalities or retrieved context.
Our experimental results demonstrate several key findings. First, we observe a consistent drop in performance for standard \MainTaskAbbr\ methods as the context becomes more reliable, which is in line with previous research~\cite{soudani2025uncertainty}. Furthermore, methods developed specifically for VLMs, such as~\cite{avestimehr2025detecting}, fail to produce meaningful results when contextual information is incorporated.
Second, we observe that finetuned methods on VLM and dataset outputs significantly outperform baseline methods, as well as methods trained for different models such as LARS~\cite{yaldiz2025lars}. This is evident in the substantial performance gap between both the finetuned LARS and our proposed method, and the base LARS model, which uses the pretrained weights provided by~\cite{yaldiz2025lars}.

Third, within the in-distribution setting, our proposed \proposedmethod\ consistently outperforms LARS. Despite incorporating additional probability tokens corresponding to unseen modality configurations (i.e., probabilities without image, without context, and without both), which could increase the data requirements, our model achieves better performance even when trained on the same amount of data. However, the increased model complexity, combined with the relatively limited training data, may explain the mixed results observed in generalization. Since LARS has fewer parameters and relies on fewer signals, it may require fewer training samples, which could explain why it outperforms \proposedmethod\ in certain transfer scenarios, such as the transition from Qwen to LLaVA.
These findings highlight that uncertainty in multimodal RAG systems is inherently complex, as it can arise at different stages of the pipeline. Methods that rely solely on image- or text-based signals, as seen in the baselines, fail to fully capture this complexity. Despite the improvements introduced by \proposedmethod, it does not consistently surpass all baseline \MainTaskAbbr\ methods under distribution shifts, indicating that robust generalization remains an open challenge.

\section*{Limitations}
Our study has several limitations. First, like other white-box uncertainty quantification methods, \proposedmethod~ requires access to token-level probability information, or logits, from the underlying VLM. In contrast, black-box UQ methods, which rely only on generated outputs, do not have this requirement and may therefore be easier to apply to proprietary or API-only models. This dependency may limit the applicability of \proposedmethod~ in settings where token probabilities are unavailable or costly to obtain.

Second, we consider only a single retrieved context. Extending \proposedmethod\ to incorporate multiple retrieved contexts may provide additional gains in accuracy of final responses.
Finally, the retrievers analyzed in this work are relatively simple. The observed differences in performance between the EVAC setting and other retrieval setups suggest that the retrieval mechanism itself could have an impact on uncertainty estimation. Exploring this effect and their interactions with \MainTaskAbbr\ methods is a promising avenue for future work.

\begin{acks}
This publication is part of the project LESSEN with project number NWA.1389.20.183 of the research program NWA ORC 2020/21 which is (partly) financed by the Dutch Research Council (NWO).
\end{acks}

\bibliographystyle{ACM-Reference-Format}
\bibliography{main}
\balance

\appendix
\clearpage
\newpage
\section{Prompt Details}~\label{app:prompt}
\begin{promptbox}
\textbf{Prompt: Generation for LLaVA 1.5-7B without context}\\
USER: <image>\\
\{question\} Answer with a single word or a short sentence.\\
ASSISTANT:
\end{promptbox}

\begin{promptbox}
\textbf{Prompt: Generation for LLaVA 1.5-7B with context}\\
USER: <image>\\
Context: \{context\}\\
\{question\} Answer with a single word or a short sentence.\\
ASSISTANT:
\end{promptbox}

\begin{promptbox}
\textbf{Prompt: Generation for Qwen3-VL-4B without context}\\
\{question\} Answer with a single word or a short sentence.
\end{promptbox}

\begin{promptbox}
\textbf{Prompt: Generation for Qwen3-VL-4B with context}\\
\{context\}\\
\{question\} Answer with a single word or a short sentence.
\end{promptbox}

\begin{promptbox}
\textbf{P(True) System Prompt}\\
You are a helpful, respectful and honest question-answer evaluator. You will be given a question, some brainstormed ideas and a generated answer. Evaluate the generated answer as true or false considering the question and brainstormed ideas. Output "The generated answer is true" or "The generated answer is false".
\end{promptbox}

\begin{promptbox}
\textbf{Prompt: P(True) without context}\\
USER: <image>\\
Question:\{question\}\\
Here are some ideas that were brainstormed:\{ideas\}\\
Generated answer:\{generated\_text\}
\end{promptbox}

\begin{promptbox}
\textbf{Prompt: P(True) with context}\\
USER: <image>\\
Context:\{context\}\\
"Question:\{question\}\\
Here are some ideas that were brainstormed:\{ideas\}\\
Generated answer:\{generated\_text\}"
\end{promptbox}

\end{document}